# First test of a cryogenic scintillation module with a CaWO$_4$ scintillator and a low-temperature photomultiplier down to 6 K

H. Kraus and V.B. Mikhailik

*Department of Physics, University of Oxford, Keble Road, Oxford OX1 3RH, UK*

**Abstract**

Future cryogenic experiments searching for rare events require reliable, efficient and robust techniques for the detection of photons at temperatures well below that to which low-temperature photomultipliers (PMT) were characterised. Motivated by this we investigated the feasibility of a low-temperature PMT for the detection of scintillation from crystalline scintillators at T = 6 K. The scintillation module was composed of a CaWO$_4$ scintillator and a low-temperature PMT D745B from ET Enterprises. The PMT responsivity was studied at T=290, 77 and 6 K using γ-quanta from $^{241}$Am (60 keV) and $^{57}$Co (122 and 136 keV) sources. We have shown that the low-temperature PMT retains its single photon counting ability even at cryogenic temperatures. At T = 6 K, the response of the PMT decreases to 51 ± 13 % and 27 ± 6 % when assessed in photon counting and pulse height mode, respectively. Due to the light yield increase of the CaWO$_4$ scintillating crystal the overall responsivity of the scintillation modules CaWO$_4$+PMT is 94 ± 15 % (photon counting) and 48 ± 8 % (pulse height) when cooling to T = 6 K. The dark count rate was found to be 20 s$^{-1}$. The energy resolution of the module remains similar to that measured at room temperature using either detection mode. It is concluded that commercially available low-temperature PMT are well suited for detection of scintillation light at cryogenic temperatures.

## 1. Introduction

Detection of photons, created by particle interaction in a scintillating volume, is an established method for determining key event parameters, such as energy, timing, or interaction type. The excellent sensitivity for single photon detection and the robustness of photomultiplier tubes (PMT) are compelling reasons for preferring PMTs for the detection of scintillation, despite impressive progress in the development of novel photodetectors [1]. Major research effort is aimed at improving the performance of PMTs and at widening the range of their applications. Such optimization is often driven by particular experimental needs; and one such example is the development of PMTs for operation at low temperatures. This development was motivated by the experimental search for dark matter based on the detection of scintillation in liquid noble gases. PMTs have been extensively tested and used in experiments at temperatures of liquid xenon (166 K), argon (87 K), nitrogen (77 K) [2, 3, 4, 5, 6, 7, 8, 9] and recently a low-temperature PMT was characterised down to a temperature of 29 K [10]. Detection of scintillation photons from crystalline scintillators at even lower temperatures is required by present and future cryogenic dark matter experiments that use a combination of heat and light signals to determine precisely the energy of an event and the type of recoil (i.e. electron or nuclear) [11, 12, 13]. Having demonstrated the feasibility of PMTs to operate at temperatures as low as that of liquid neon, the natural extension is to evaluate the performance of PMTs at temperatures close to that of liquid helium. Here we report first results of the characterisation of a cryogenic scintillation module composed of a $CaWO_4$ scintillator and low-temperature PMT (Electron Tubes (UK), model D745) inside a $^4$He cryostat.

## 2. Experiment

The PMT studied in this work is a low-temperature 2 inch diameter bialkali model D745B, produced by ET Enterprises (UK). The specifications of the PMT are listed in Table 1. The scintillating $CaWO_4$ crystal of diameter 10 mm and height 5 mm was produced by SRC Carat (Lviv, Ukraine). Fig. 1 shows the design of the cryogenic scintillation module. The PMT (1) and crystal (2) are encapsulated in a vacuum-tight enclosure made of copper. The enclosure is attached to a long tube from which it can be suspended and lowered into the liquid helium inside a $^4$He cryostat. The long tube is also used as pumping line and it contains the cabling connecting the room temperature electronics to the PMT and the temperature sensors.

HERE TABLE 1

The scintillator crystal within the enclosure is surrounded by a reflector made of a 3M foil (3), exhibiting reflectivity close to 100% throughout the visible wavelength range [14]. The gap between the scintillator and the PMT photocathode was ~2 mm. Scintillation is stimulated by γ-radiation from $^{241}$Am (60 keV, 33 kBq) or $^{57}$Co (122 and 136 keV, 72 kBq) sources (4) placed 5 mm away from the crystal. To avoid pile-up of individual scintillation events, the count rate must be adjusted. The risk of pile-up increases with cooling due to the principal time constant of $CaWO_4$ increasing from 9.2 μs at 295 K to 360 μs at 4.2 K [15]. To reduce pile-up to a negligible level at low temperature, the event rate was adjusted to less than ~100 Hz at room temperature by inserting copper sheets (2 mm for the $^{241}$Am source and 5 mm for the $^{57}$Co source) between source and scintillator crystal.

To ensure cooling of the PMT's glass body, it is thermally connected to a copper base using a sleeve made of 10×0.1 mm$^2$ copper strips. The PMT temperature was monitored by two Lakeshore DT-470 Si-diodes, pressed against the long side of the glass body, very near to the photocathode (5). With the temperature dependence of the scintillation decay constants of $CaWO_4$ well documented [15], the temperature of the scintillating crystal can be derived from the value of the decay time constants obtained during the measurements. The data show that the temperature of the crystal was very close to the expected 290 K, 77 K and 4.2 K when the module was characterised at room temperature, or the temperatures of liquid nitrogen or liquid helium, respectively.

The PMTs used in this study were positively biased, using a linear voltage divider network with 330 kΩ resistors (6). The voltage divider was placed in vacuum at the room-temperature part of the pumping tube and connected to the PMT, using Manganin wires. This prevented the conduction of the heat generated by the resistive components of the voltage divider to the cold PMT. To provide thermal heat sinking for the internal parts of the PMT, the pins connected to the dynodes were potted in Stycast 2850 FT that fills the copper base (8). When the module was immersed into liquid helium, the temperature of the PMT with an applied voltage of 1830 V remained constant at 6.0±0.2 K.

PMTs should not be exposed to a helium atmosphere as this deteriorates the vacuum of the PMT due to the high diffusivity of helium in glass. Thus, the container housing the PMT was evacuated to a pressure of $< 2\times10^{-4}$ mbar before immersing the setup into a cryogenic liquid.

During the experiment with the setup immersed in a cryogenic liquid, the pressure measured at the room temperature end of the pumping tube was $<10^{-5}$ mbar due to cryo-pumping. Since no exchange gas was used, in order to ensure equilibrium across the device, the enclosure was kept in liquid nitrogen and liquid helium overnight before beginning the measurements.

To measure the scintillation response of the PMT, we used the photon counting technique developed by us and intensively used for the investigation of the temperature dependence of scintillation properties of materials [16]. A detailed description of the original experimental setup can be found in reference [17] and subsequent papers [15, 18], which discuss various improvements to the technique. For this experiment we designed a fast preamplifier (bandwidth 100 MHz, amplification 10 dB) to ensure sufficient signal-to-noise ratio without loss of single photon resolution and allowing matching of the impedances of PMT and transition line [19]. Single photon counting allows determining the energy of particle-induced scintillation events from distributions of the single photon signals (SPSs), detected by the PMT (see section 3.1). This allows analysing the scintillation process when both, decay time of the scintillation and the amplitude of the detected SPSs vary significantly due to changes with temperature in the response of the scintillator and the PMT. Measurements were carried out in which ~1500 signal traces (scintillation events from $CaWO_4$) were recorded at fixed temperatures: 290, 77 and 6 K.

## 3. Results

### 3.1 Time profile of scintillation events

Figure 2 shows typical traces of α- and γ-scintillation events, exited by a $^{241}$Am source in $CaWO_4$ at room temperature and detected by the D745B PMT. The figure shows that single photon signals (SPS) from a low-energy γ-scintillation event (60 keV) can be identified clearly. Conversely, high-energy α-particles (5.5 MeV) produce bursts of SPSs in a relatively short (a few tens of µs) time interval and the time profile of the scintillation event exhibits a pronounced pile-up of SPSs during the initial part of the event. Such a high rate of SPS is a complication that is best to avoid as it can affect the determination of energy when using the single photon counting mode.

The scintillation detectors used in cryogenic searches for rare events operate at low temperatures when decay time constants can be two orders of magnitude longer compared with the room temperature values [15, 16]. Consequently, the rate of photons produced during scintillation events at cryogenic temperature is two orders of magnitude lower, while the increase in the light yield of the scintillator is much smaller and hence pile-up of SPSs is not an issue. It is easy to distinguish SPS using a sampling interval of ~10 nsec and electronics with $>\sim100$ MHz bandwidth. Fig. 3 shows two scintillation events (122 keV γ-quanta from $^{57}$Co) recorded by the PMT under test, one for room temperature (trace 1) and one for low temperature (trace 2). This figure shows the distinctive differences of the scintillation events in duration and amplitude; the former caused by the temperature dependence of the scintillation time constant and the latter by the PMT. Inspection of the SPSs at low temperature showed that their average maximum value decreases compared with that of room-temperature events. The amplitude of SPS is determined by the amplification process that relies on the effect of secondary electron emission by the dynodes. This observation suggests that at low temperature the dynodes in the PMT exhibit lower emissivity. Since the pulse height spectrum detected by a PMT is proportional to the total charge collected by the anode, a change of the PMT gain will inevitably affect the response characteristic of the device, as has been documented in a number of previous studies [4, 6, 7, 8, 10, 10] (see also discussion in section 4 of this paper). Identifying SPS in photon counting mode is less dependant on the charge that each individual photon contributes as we merely aim to recognize photons and then count them. The temperature variation of this parameter is mainly dependant on the change of quantum efficiency of photocathode. This makes photon counting an attractive option for the measurements at low temperatures.

*3.2. Temperature dependence of scintillation response of the scintillation module*

The performance of the scintillation module was studied in photon counting mode using 60 keV ($^{241}$Am) and 122 keV ($^{57}$Co) γ-quanta. The effect of temperature on the distribution of the number of SPS per scintillation event from $CaWO_4$, irradiated with γ-quanta from $^{241}$Am and $^{57}$Co is displayed in Fig. 4. The results of these measurements are summarised in the upper part of Table 2 (photon counting). The response (weighted mean from 60 and 122 keV excitation) of the scintillator module ($CaWO_4$ crystal and the D745B PMT) is 130 ± 19% (at T = 77 K) and 94 ±15 % (at T = 6K), compared with the response at T=290 K.

HERE TABLE 2

The responsivity of the PMT alone was obtained by correcting for the increase of scintillation yield of $CaWO_4$ with cooling. The change with temperature of the scintillation light yield of $CaWO_4$ has been well documented [15, 16, 17]. Here, we use a crystal cut from the same boule as the sample whose light yield and decay kinetics have previously been studied in the temperature range 0.02 – 350 K [15]. Based on these results the light yield of the $CaWO_4$ scintillator has been set to unity at T=290 K, 1.6±0.3 at T=77 K and 1.8±0.3 at T=6 K. This allows taking into account effects introduced by thermal changes of the scintillator characteristics and filtering out the contributions caused by the PMT alone. Following this correction it was found that the relative response of the PMT decreased by ~20 % when cooled to T = 77 K. At the temperature of liquid helium the detection efficiency of the PMT is 51 ± 13 % of its room temperature value. Though the statistical error of this estimate is fairly large, these results show that the low-temperature PMT is well suited for detection of scintillation light at cryogenic temperatures in the photon counting mode.

The variation of the PMT responsivity with temperature can be evaluated using pulse height spectra, obtained from the single photoelectron response [6, 10, 10, 20] or from the response due to excitation by a radiation source [2, 4, 8, 21]. This is because analogue integration of the PMT signal was and still remains a simple, affordable and hence popular method of signal processing. It is generally accepted that the amplitude of the integrated PMT signal is proportional to the total charge collected by the photocathode and to first approximation is proportional to the number of detected photons [22]. Therefore, both the pulse height spectrum and the distribution of SPS provide a measure for the PMT responsivity. To allow comparison with earlier studies, we have also performed analysis of our results by integrating each signal trace and plotting the respective histogram that is analogous to the pulse height spectrum.

Fig. 5 shows a typical set of pulse height spectra for the excitation with γ-quanta from $^{241}$Am (60 keV) and $^{57}$Co (122 keV), detected by the scintillation module at different temperatures. The results are listed in the lower part of Table 2 (pulse height). The average change in the responsivity of the scintillation module $CaWO_4$+PMT when cooled to T = 77 K is essentially the same as obtained in photon counting mode (129 ± 18 %). The increase is merely due to the higher light yield of $CaWO_4$ at lower temperature. Taking into account the variation with temperature of the light yield of $CaWO_4$, the response of the PMT at T = 77 K was found to be 81 ± 23 % that at 295 K. The pulse height spectrum of γ-induced scintillation detected by the PMT (T = 6 K) exhibits an even more significant reduction of the responsivity due to the reduction of charge collected by the PMT. The observed decrease of the relative response of the module (48 ± 8 %) corresponds to a decrease of the PMT responsivity to 27 ± 6 % of its room temperature value. Given that the number of produced photoelectrons changes less significantly (by only a factor two) it is most likely that a decrease of the secondary electron emission of the dynodes is significantly contributing to the deterioration of the PMT performance. Therefore the operation of the scintillation module in the pulse height mode at very low temperatures, although possible, seems to be inferior when compared with operation in the photon counting mode.

Regarding energy resolution, to first approximation, the increase of detector response facilitates improvement of this parameter. Indeed, Table 2 shows a consistent improvement of the energy resolution with an increase of the scintillation module response from 290 K down to 77 K. In line with such dependency this parameter degrades when the responsivity of the scintillation module decreases with further cooling. Fortuitously, the decrease of the PMT response is mostly cancelled by the concomitant increase of light yield of $CaWO_4$. At T=6 K the energy resolution of the scintillation module remains very similar to that at room temperature.

*3.3. Dark counts*

The dark count rate is a concern for measurements of low intensity signals, especially at cryogenic temperatures when a long (millisecond) recording time is needed to fully capture a scintillation event. To study this characteristic of the PMT, the radioactive sources and the $CaWO_4$ scintillator were removed from the module. The PMT was kept in the evacuated module for more than 12 hours before measurements to ensure a drop of the dark count rate induced by exposure of the photocathode to ambient light. About 600 traces of the PMT output signals were recorded, each lasting 2.6 ms (duration of contiguous recording used for low-temperature measurements). The total number of detected PMT pulses was counted and divided per total acquisition time to obtain the dark count rate. Figure 6 shows the dark count rate as function of anode voltages at different temperatures measured for the ET Enterprise D745B PMT. The dark count rate measured at the same anode voltage decreases by a factor of 2.5 with cooling of the PMT to 18 K; at this temperature, and an anode voltage of 1830 V, the typical dark count rate is 20 $s^{-1}$.

The dark count rate is predicted to reduce with cooling due to a decrease of thermally induced emission of electrons from the photocathode in the absence of optical stimulation [23]. However this effect is important merely for temperatures of >100 K since the process requires relatively high activation energies. At low temperatures the thermal generation is quenched and we expect that the main contributors to the dark count rate are radioactivity and cosmic rays that produce spurious events in the PMT and surrounding [24]. We would like to add that available, up-to-date experimental results on the temperature dependence of dark counts are inconclusive. A decrease of dark counts has been observed for two types of PMTs when cooling to the temperature of liquid argon [9], which is consistent with our finding. There are however other studies that report an increase of the dark count rate with cooling to temperatures of liquid nitrogen [6], liquid neon [10] and liquid helium [25]. More investigation will have to be carried out to separate the various effects contributing to an overall increase of decrease of the dark count rate when cooling. Contributions to the behaviour observed could arise from the particular PMTs being tested, the method, or other, so far unidentified causes.

### 4. Discussion

The variation of the PMT response with temperature is a complex function of several temperature dependant parameters, i.e. photocathode sensitivity, dynode multiplication, window transparency, etc., that are difficult to separate. Earlier studies on standard PMTs have shown that the temperature dependence of the PMT sensitivity is dominated by an increase in the electrical resistivity of the photocathode when cooled. This causes a gradual build-up of electrical charge within the photocathode that eventually leads to a loss in sensitivity [3, 4, 5, 8]. Therefore, the photocathodes of PMT for low-temperature applications are produced with conducting strips or a nearly transparent conducting layer of a metal to reduce this charging. Either of these designs, however, is on the expense of decreased quantum efficiency. Thus, the quantum efficiency for a PMT with platinum underlay decreases compared with that of a standard one [9]. Obviously, this is a limiting factor common to all PMTs and finding a way around the problem of photocathode charge-up, but without significant loss of quantum efficiency, is a most important issue that needs addressing.

There are several recent studies of low-temperature PMT that report both increase [2] and decrease [6, 8] of the PMT response down to liquid nitrogen temperature. Very recently a low-temperature Hamamatsu R5912MOD PMT with bialkali photocathode and platinum underlay was characterised down to T = 29 K by measuring the single photoelectron response [10]. It was found that the PMT gain exhibits a gradual decrease with cooling from room temperature down to 75 K. Below this temperature the authors reported a dramatic drop of the PMT gain with cooling (almost two orders of magnitude). It should be noted that this result is based on using the pulse height spectrum produced by the PMT under test. Thermally induced deformations of the electron optics, as well as changes with temperature of the conductive and emissive characteristics of the dynode coating were suggested as possible reasons for such dramatic changes. This finding differs substantially from our results: the investigated PMT showed only a moderate (by a factor of four) decrease of the responsivity with cooling when measured in the pulse height mode. In principle this effect can be counteracted by increasing the anode voltage [6], but the drawback is an increase of dark counts. The PMT under investigation maintains better performance in photon counting mode when cooled down to cryogenic temperatures. It should be noted that this is consistent with the previous study [10] where the photon detection efficiency of low-temperature bialkali PMT evaluated in photon counting mode has been shown to change very little with cooling. We observed a responsivity reduction by factor of two and it is likely that this decrease is partially due to the reduction of the PMT gain since lower signal-to-noise ratio affects the identification of SPS. This observation leads us to an important conjecture: the quantum efficiency of bialkali photocathode in the range of emission of the $CaWO_4$ scintillator is only slightly affected by cooling to cryogenic temperature.

According to the results of recent studies, the spectral sensitivity of PMT exhibit two trends with temperature. It increases with cooling in the short wavelength range [2, 9, 26], which is explained by the reduction of lattice scattering of photoelectrons driven towards the surface of the photocathode [2, 9]. Conversely, the sensitivity of a bialkali photocathode decreases in the red [4, 7, 9, 23] due to a change of the energy band structure of the photocathode material. Since the emission band of the CaWO$_4$ scintillator with a maximum at 420 nm overlaps with the central part of the quantum efficiency curve of the bialkali photocathode neither the long wavelength cut-off nor the short wavelength enhancement feature prominently as a temperature effect. Subsequently, due to the good match of the spectral properties of the CaWO$_4$ scintillator and the bialkali photocathode, the cryogenic module exhibits only a modest change in the photon detection efficiency as a function of temperature.

### 5. Conclusion

The development of light detectors for operation in a cryogenic environment has stimulated our interest in investigating the feasibility of a PMT readout for low-temperature scintillators. We carried out a first characterisation of a scintillation module composed of a CaWO$_4$ scintillator and a low-temperature PMT (ET Enterprises, model D745B) in a $^4$He cryostat down to T = 6 K. The use of a photocathode with a conductive underlay in a low-temperature PMT helps with preventing the undesired accumulation of charge, thereby making possible operation of this PMT at cryogenic temperatures.

Measurements of the number of photons per scintillation event excited by ionising radiation showed that the PMT operates adequately in the photon counting mode. In detecting scintillation events from CaWO$_4$ at T=6 K the relative response of the PMT exhibits a ~50 % decrease compared to the room temperature value. Nonetheless the response of the scintillation module CaWO$_4$+ PMT at T = 6 K is very close to that at room temperature (94 ± 15 %). Cooling the PMTs to cryogenic temperatures also substantially reduces the dark count rate to 20 s$^{-1}$ (anode voltage 1830 V).

The temperature dependence of the PMT responsivity was also analysed using the pulse height spectrum of scintillation events. It is found that the relative response evaluated in the pulse height mode is more affected by temperature because of the change of the PMT gain that influences the total charge collected. The relative response of the PMT in pulse height mode at T=6 K is 27 ± 6 % of its room temperature value.

Finally it has been shown that despite of the observed reduction of the PMT response, the energy resolution of the scintillation module measured in either photon counting or pulse height mode at T=6 K is very similar to that at room temperature. It is obvious that the low-temperature PMT exhibits promising characteristics at temperatures close to that of liquid helium. Such PMT could be well suited for the detection of photons from cryogenic scintillators operating at even lower temperatures.

### Acknowledgment

The study was supported in part by the Science and Technology Facilities Council (STFC). Authors are indebted to Mr. M. Tacon and Mr. A. Grace for their continuous efforts in building the cryogenic scintillation module.

Table 1. Specification and settings of the PMT under test

| Parameter | D745B |
|---|---|
| Photocathode | K-Cs-Sb + Pt underlay |
| Window | Borosilicate glass |
| Diameter, inch | 2 |
| Gain | $7 \times 10^6$ |
| Dark current, nA | 1 |
| Dynodes | 12 linear focused |
| Dynode material | SbCs |
| Voltage divider | 5R-12×1R (R=330 kΩ) |
| Anode voltage applied (maximum), V | 1830 (2000) |

Table 2. Response characteristics of the scintillation module composed of a $CaWO_4$ scintillator and D745B PMT. Excitation with $^{241}$Am (60 keV) and $^{57}$Co (122 keV), anode voltage 1830 V.

| T, K | Peak position | | Energy resolution FWHM , % | | $CaWO_4$ light output | Relative response of module | | Relative response of PMT | |
|---|---|---|---|---|---|---|---|---|---|
| | $^{241}$Am | $^{57}$Co | $^{241}$Am | $^{57}$Co | | $^{241}$Am | $^{57}$Co | $^{241}$Am | $^{57}$Co |
| Photon counting | | | | | | | | | |
| 290 | 57±10 | 98±16 | 40 | 38 | 1 | 1 | 1 | 1 | 1 |
| 77 | 70±8 | 137±18 | 32 | 30 | 1.6±0.3 | 1.22±0.26 | 1.40±0.29 | 0.76±0.30 | 0.87±0.36 |
| 6 | 50±9 | 99±14 | 42 | 33 | 1.8±0.3 | 0.87±0.22 | 1.01±0.21 | 0.48±0.12 | 0.56±0.21 |
| Pulse height | | | | | | | | | |
| 290 | 1.35±0.24 | 2.89±0.46 | 43 | 37 | 1 | 1 | 1 | 1 | 1 |
| 77 | 1.86±0.24 | 3.63±0.35 | 31 | 23 | 1.6±0.3 | 1.37±0.30 | 1.25±0.23 | 0.85±0.36 | 0.78±0.29 |
| 6 | 0.69±0.13 | 1.37±0.19 | 45 | 33 | 1.8±0.3 | 0.51±0.13 | 0.47±0.10 | 0.28±0.09 | 0.26±0.08 |

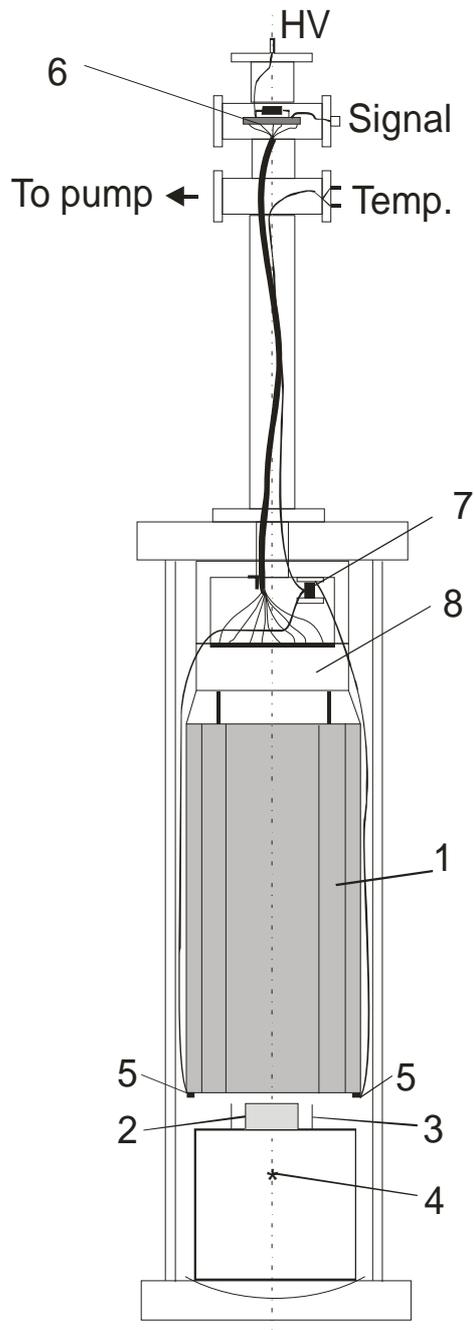

Fig. 1: Experimental setup: 1 – PMT, 2 – CaWO$_4$ scintillator, 3 – reflector, 4 – radioactive source, 5 – temperature sensors, 6 – voltage divider, 7 – heat sink, 8 – copper base with PMT pins potted in Stycast 2850 FT.

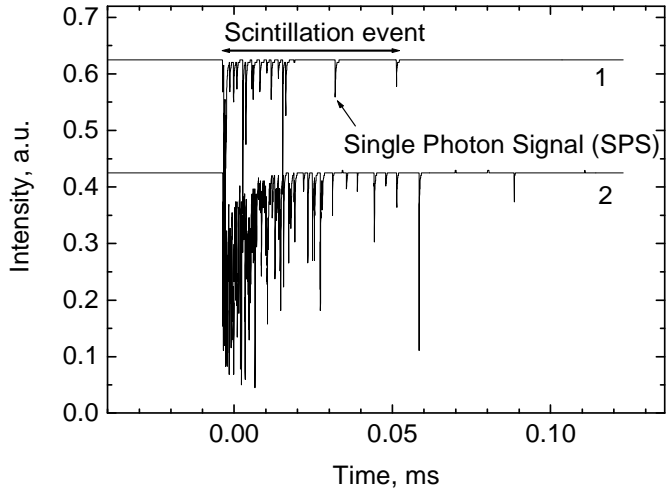

Fig. 2: Typical signal traces of scintillation events detected by the ET Enterprises D745B PMT at T = 290 K. Excitation with 60 keV γ-quanta (1) and 5.5 MeV α-particles (2) from a $^{241}$Am source. The figure also shows a single photon signal (SPS) and scintillation event as they are referred to throughout this paper.

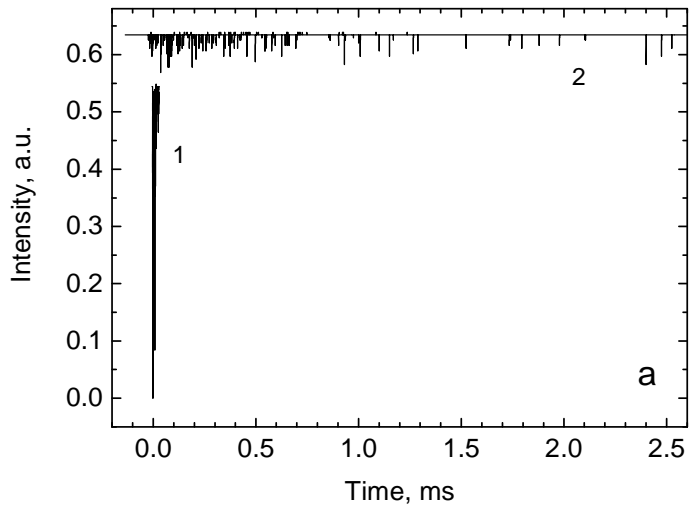

Fig.3: Comparison of typical signal traces of scintillation events detected by the PMT at T = 290 (1) and 6 K (2). Excitation with 122 KeV γ-quanta from a $^{57}$Co source.

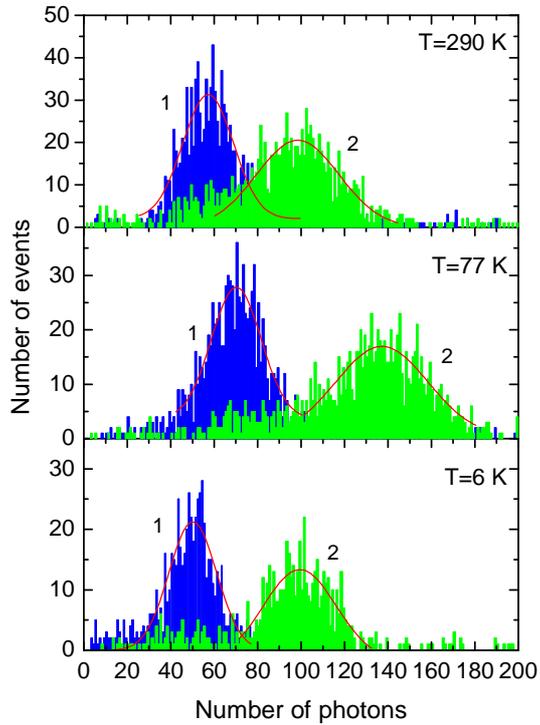

Fig. 4: Distributions of the number of single photon signals (x-axis: number of photons) per scintillation event detected by the module at T = 290 K, 77 K and 6 K. Excitation with γ-quanta from $^{241}$Am (1) and $^{57}$Co (2) sources, anode voltage 1830 V. The spectra are fitted by single Gaussians (red). The hump in the low energy part of the $^{57}$Co spectra is due to the escape of X-rays excited in K-shell of tungsten.

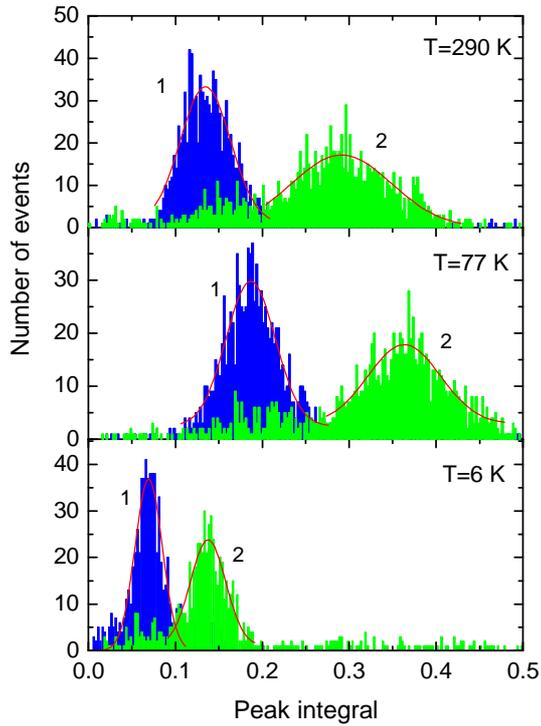

Fig. 5: Pulse height distributions, derived from PMT signal traces. For each scintillation event, the total area under all SPSs was calculated. Data were taken at temperatures of 290, 77 and 6 K. Excitation with γ-quanta from $^{241}$Am (1) and $^{57}$Co (2) sources, anode voltage 1830 V. The spectra are fitted by single Gaussians (red). The hump in the low energy part of $^{57}$Co spectra is due to the escape of X-rays excited in K-shell of tungsten.

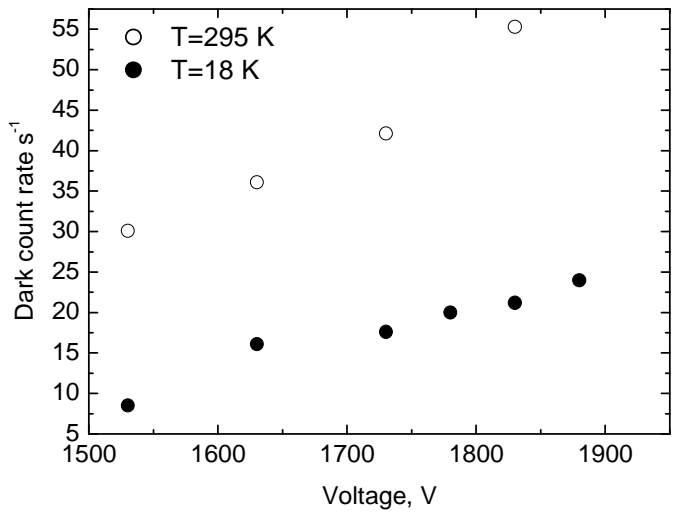

Fig. 6. Dark count rate of the ET Enterprises D745B PMT as function of the anode voltage at T=295 and 18 K.